\documentclass[conference]{IEEEtran}
\IEEEoverridecommandlockouts

\usepackage{subcaption}
\usepackage{tikz}
\usetikzlibrary{arrows}
\usepackage{verbatim}
\usepackage{pgfplots, pgfplotstable}
\usepackage{url}
\usepackage{tabularx}
\usepackage{balance}
\usepackage{booktabs}
\usepackage{float}
\usepackage{cite}
\usepackage{amsmath,amssymb,amsfonts}
\usepackage{algorithmic}
\usepackage{graphicx}
\usepackage{textcomp}
\usepackage{xcolor}
\pgfplotsset{compat=1.7}
\usepackage{wrapfig}
\usepackage{enumitem}

\begin{document}

\title{Towards a Zero-Trust Micro-segmentation Network Security Strategy: An Evaluation Framework}

\author{\IEEEauthorblockN{Nardine Basta}
\IEEEauthorblockA{\textit{Macquarie University}\\
nardine.basta@mq.edu.au}
\and
\IEEEauthorblockN{Muhammad Ikram}
\IEEEauthorblockA{\textit{Macquarie University}\\
muhammad.ikram@mq.edu.au}
\and
\IEEEauthorblockN{Mohamed Ali Kaafar}
\IEEEauthorblockA{\textit{Macquarie University}\\
dali.kaafar@mq.edu.au}
\and
\IEEEauthorblockN{Andy Walker}
\IEEEauthorblockA{\textit{ditno} \\
Andy@ditno.com}

}

\maketitle

\begin{abstract}
Micro-segmentation is an emerging security technique that separates physical networks into isolated logical micro-segments (workloads). By tying fine-grained security policies to individual workloads, it limits the attacker’s ability to move laterally through the network, even after infiltrating the perimeter defences. While micro-segmentation is proved to be effective for shrinking enterprise networks attack surface, its impact assessment is almost absent in the literature. This research is dedicated to developing an analytical framework to characterise and quantify the effectiveness of micro-segmentation on enhancing networks security. We rely on a twofold graph-feature based framework of the network connectivity and attack graphs to evaluate the network exposure and robustness, respectively. While the former assesses the network assets connectedness, reachability and centrality, the latter depicts the ability of the network to resist goal-oriented attackers. Tracking the variations of formulated metrics values post the deployment of micro-segmentation reveals exposure reduction and robustness improvement in the range of 60\% -- 90\%.
\end{abstract}

\section{Introduction}
Micro-segmentation \cite{ACSC} is a core component of the zero-trust security concept. It creates secure zones across cloud and data centre environments to isolate the different application workloads and secure them independently. It further generates dynamic access control policies that limit network and application flows between workloads. Accordingly, it protects the network assets and provides control and visibility over the growing amount of east-west traffic across the organization which bypasses the traditional firewalls.

Autonomously modelling application behaviour and accounting for workloads is a major challenge towards achieving a zero-trust architecture. This can be attributed to the fact that enterprises data centre has evolved from on-premises infrastructure to a distributed facility with inter-connected cloud infrastructure where networks, applications and workloads are virtualized in multiple private and public clouds. Hence, in addition to the introduced complexity of the network architecture, enterprises have little confidence in their underlying network structure and connectivity.

While several micro-segmentation solutions are currently offered by industry (e.g., {\it ditno} \cite{ditno}, Cisco \cite{Cisco}, and others), the general question of how effective and efficient these security controls are, still persists. In fact, little is known about how implementing these controls would compare to security risks within flat networks. This is not only essential to understand what level of protection is offered by the different security controls, but is also important to justify resources and investment to augment existing controls. 

In this paper, we leverage attack-graph generation and probabilistic reasoning framework for comprehensive security and effectiveness analysis of network micro-segmentation. The main contributions of our work are as follows: 

\begin{itemize}[noitemsep,topsep=0pt]
\item We develop a framework to assess and quantify the effectiveness of micro-segmentation in reducing the enterprise network assets risk of exposure to insider and outsider threats. We further analyse the robustness of the network by measuring its ability to resist goal-oriented attackers. 

\item To generate a reproducible and objective evaluation framework we base our metrics on graph feature analysis of the network connectivity and attack graphs.

\item We rely on two enterprises network data to perform an empirical analysis of the effectiveness of micro-segmentation in light of the formulated framework. To the best of our knowledge, no previous work to quantify the impact of micro-segmentation, while relying on real-life network traffic, exists.

\item Using data-sets from two real-world enterprise networks, we show that micro-segmentation successfully doubles the chain an attacker is forced to pursue to compromise a target network asset by automatically identifying and blocking the illegitimate network internal connections. 

\item While micro-segmentation is unable to reduce the network vulnerabilities, we show that modifying the system security configurations influences the likelihood of exploiting the vulnerabilities. We demonstrate that micro-segmentation decreases the network misconfigurations by 65\% and the number of possibilities an intruder can exploit the network vulnerabilities by 99\%. 

\item Additionally, we show that micro-segmentation contributes to enhancing the visibility of the network architecture. By identifying and classifying the network applications/services workloads, it enables highlighting the network misconfigurations and illegitimate connections.

\item Our centrality metrics provide insight into the network weak links that should be prioritised for redemption.

 %
%\item We deploy the proposed networks into different enterprises and analyse the formulated metrics {\it prior} to and {\it post} the deployment of micro-segmentation. We find that that 
%micro-segmentation is one of the most effective strategies to protect against cyber threats reducing the overall risks by $\leq$ 90\%. In particular, we find that our deployed framework improve the network robustness by 99\%. 

% \item Wemicro-segmenting the network of Enterprise A, reveals the efficacy of micro-segmentation to significantly reduce the total number of attack paths by 99\%

% \item {\textbf{DK: We need to illustrate some of the findings from the evaluation Framework here. What are the main take aways? We have one of them in the abstract. We need to provide a few numbers from the Analysis and Results section here. }}

% \textbf{DK: More generally, in the Analysis and Results section we need to Clearly identify the main Results/conclusions from the experimental analysis. This is a priority at this stage. } 

\end{itemize}

%The rest of the paper is organised as follows: Section \ref{sec:bacckground} presents background information about micro-segmentation and overviews related work. Section \ref{sec:framework} introduces the characterization framework where Subsection \ref{sec:expo} discusses the exposure metrics while Subsection \ref{sec:Risk} proposes the robustness metrics. In Section \ref{sec:aresults} we present the results and an empirical analysis of the different formulated metrics and conclude the paper in Section \ref{sec:conc}.

\section{Background and Related Work}\label{sec:bacckground}

\subsection{Background}\label{subsec:bacckground}

\begin{figure}[ht]
  \subfloat[Flat network]{
	\begin{minipage}[c][1\width]{
	   0.23\textwidth}
	   \centering
	   \includegraphics[width=1\textwidth]{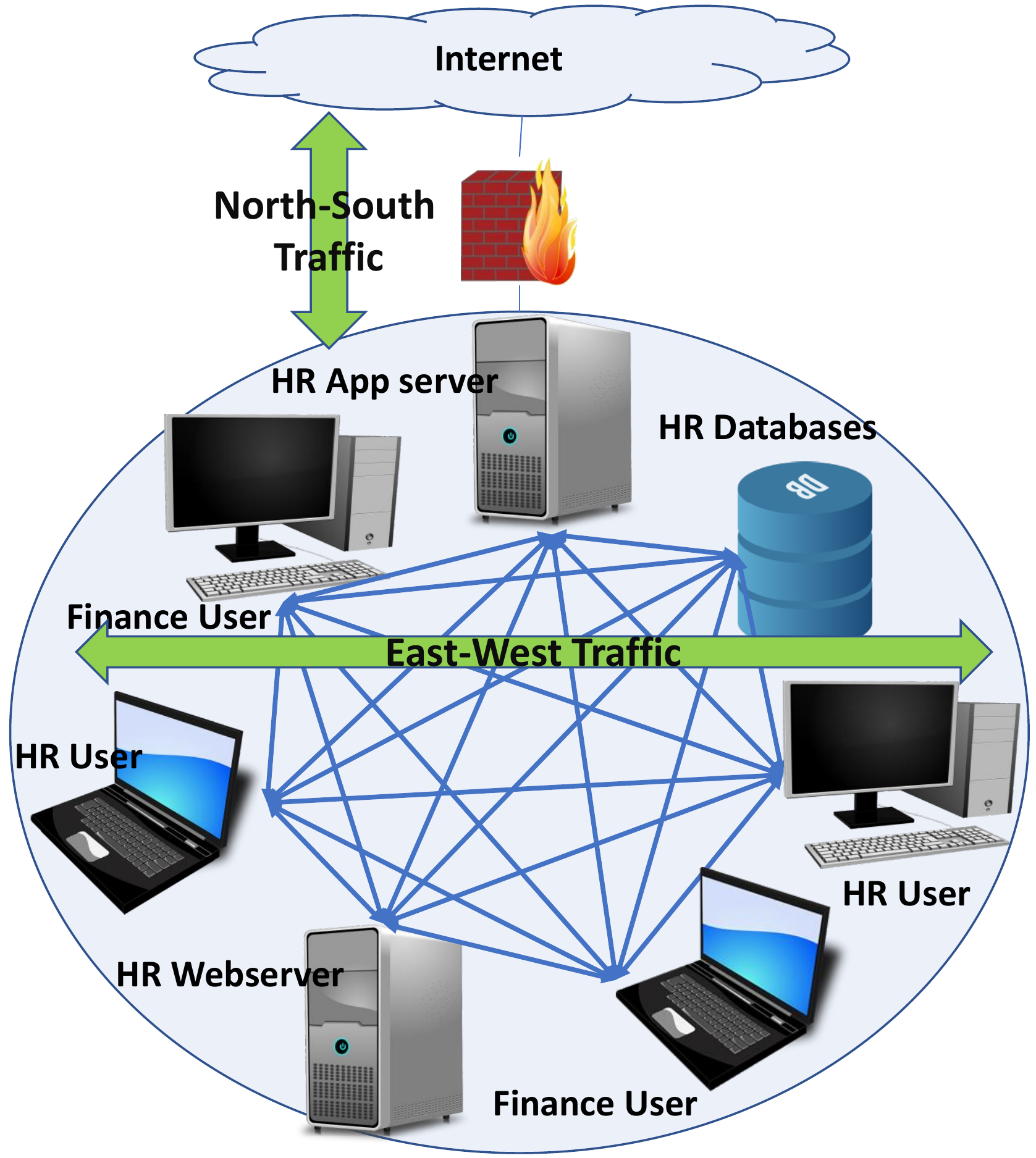}
	\end{minipage}
	\label{fig:flat}
	}
 %\hfill 	
  \subfloat[Segmented network]{
	\begin{minipage}[c][1\width]{
	   0.23\textwidth}
	   \centering
	   \includegraphics[width=1\textwidth]{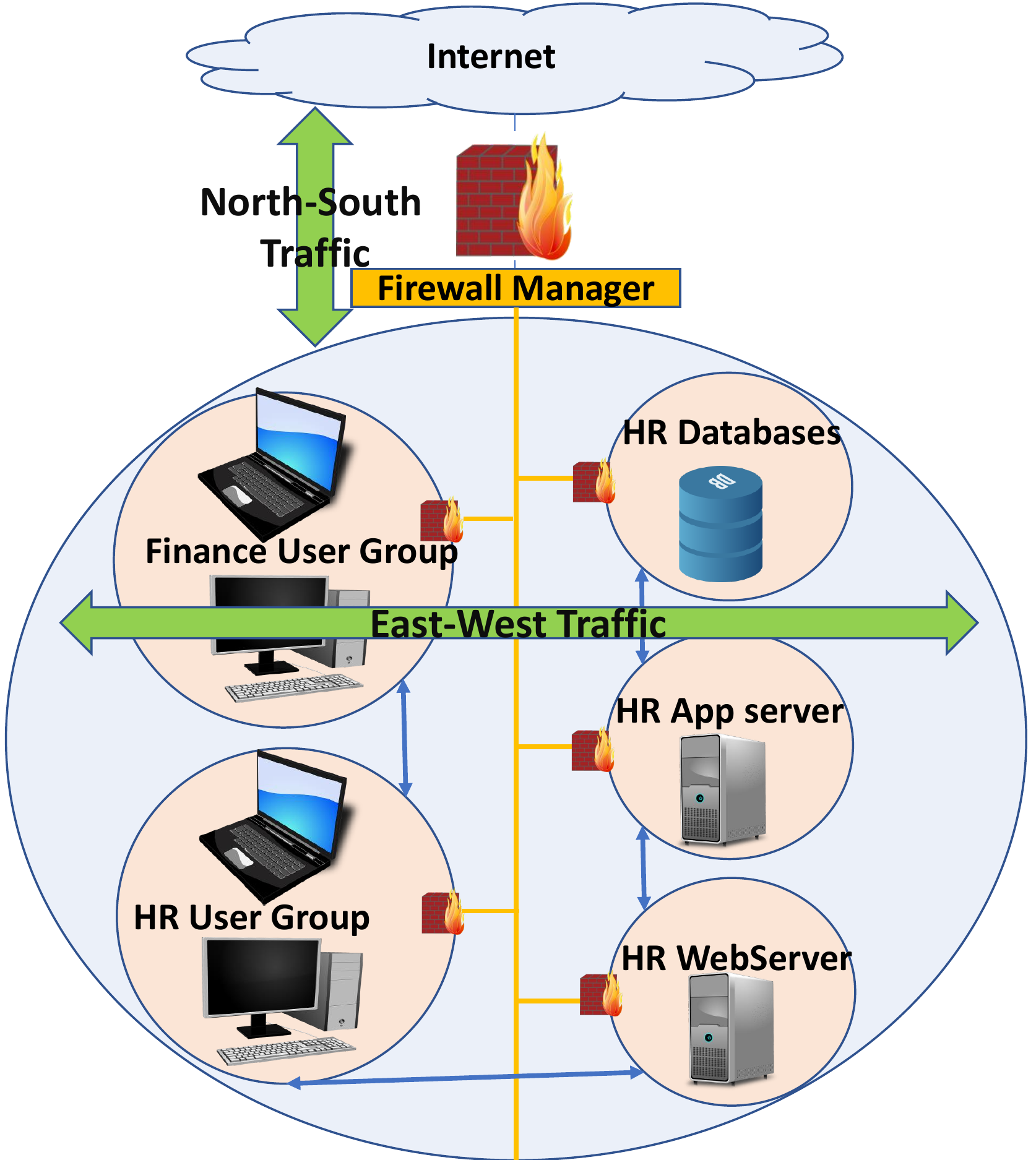}
	\end{minipage}
	\label{fig:seg}
	}

\caption{An example of flat and micro-segmented network topologies of an enterprise network. In Figure~\ref{fig:flat}, the network assets are fully connected with no restrictions on the internal communication. In Figure~\ref{fig:seg}, the network is micro-segmented into different workloads (shown in circles), thus restricting the east-west traffic.}
\label{fig:flat_seg}
%\vspace{-0.3cm}
\end{figure}

Micro-segmentation is an implementation of a distributed virtual firewall that regulates access to network assets based on security rules that have been determined on each workload (micro-segment). The firewalls examine the internal network traffic up to layer four (transport layer) of the Open Systems Interconnection (OSI) model and enforce access control through the generated micro-segmentation policies. Figure \ref{fig:flat_seg} depicts the connectivity structure of a sample enterprise network before and after micro-segmentation.

Unlike traditional firewalls that have no business logic or context into applications, the micro-segmentation firewalls are often implemented in software and deployed on each network workload. Machine-learning techniques are used to leverage the meta-data from micro-segments for firewall rules generation and generalisation across similar workloads. The network micro-segmentation firewalls are all linked to central management that pushes the policies to each firewall. It, therefore, enables granular policy enforcement throughout the enterprise network, not just at the perimeter.

There exist different possible architectures for dividing IT services into multiple tiers. Services are most commonly divided into three tiers; web servers, application servers and databases. Nevertheless, some approaches combine the web server and application server tiers. Workloads within a single IT service are allowed to communicate to each other through very restrictive rules to minimise the propensity of lateral movements. As a result, micro-segmentation offers better visibility and accountability of network resources and communication among micro-segments.

\subsection{Related Work}\label{subsec:rwork}

Efforts have been deployed in the domain of measuring and quantifying network security in general and the effectiveness of firewalls in particular. In fact, quantifying security is one of the well-recognized open problems \cite{quant1, quant2, quant3} where most of the approaches quantify the overall network security performance \cite{quant3, quant4, quant5, quant6, quant7}. Some approaches are specially dedicated to analysing the deployed firewalls \cite{quantF3} and quantifying their effectiveness \cite{quantF1, quantF2, quantF4}. 

Although micro-segmentation has gained momentum in reducing the network attack surface, improving breach containment and strengthening regulatory compliance, the assessment and quantification of its performance are almost absent in the academic and industrial literature. Only one approach is found to quantify the overall efficacy of micro-segmentation by measuring the time it takes the attacker to traverse the network and compromise its assets \cite{illum}. 
Nevertheless, the utilised metric is based on the attacker's skills rather than the strength of the micro-segmentation security controls. Additionally, the tests were conducted in a simulated network environment rather than a real network. 

In this paper, we fill the gap by providing more comprehensive and reproducible metrics to accurately represent the impact of micro-segmentation on enhancing the security of enterprise networks. 

\section{Micro-Segmentation Impact Quantification Framework}\label{sec:framework}

Next, we outline our framework which assesses the impact of micro-segmentation on enterprises network security through quantitative metrics based on graph analysis. 
In particular, we characterise the impact of micro-segmentation on the network {\it exposure} and {\it robustness}. 

\subsection{Network Exposure}\label{sec:expo}
Keeping pace with the dynamic nature of the industry, enterprise networks are evolving fast. In fact, network exposure, due to expanding the network connectivity, is the new IT risk many businesses are ignoring at their peril. Hence, comes the importance of micro-segmentation to block the illegitimate, insecure or unneeded connections and control the east-west traffic within the enterprise network.  

%\textbf{Problem formulation:} 
To formally define network exposure, let $\mathit{C(A,V,w)}$ be the network connectivity directed graph where $\mathit{V}$ is the set of graph vertices representing the network assets. $\mathit{A}$ is the set of graph directed edges indicating that the connected vertices are allowed to communicate, where $\mathit{A \subseteq \{ (x,y) \;| \; (x,y) \in V^2 \land x \neq y\}}$. $\mathit{w: A \rightarrow \mathbb{R}}$ is a weight function representing the number of services that the network assets are allowed to use for communicating in the direction of the connection. We classify the network exposure metrics into three categories: {\it connectedness}, {\it reachability}, and {\it centrality}.

\noindent\textbf{Connectedness: }A direct indicator of the internal network exposure is the amount of connections allowed to and from the different enterprise network assets. In fact, the higher the connectedness between the network assets, the more the possibilities/ways that are presented to attackers in order to achieve their malicious target.   

We formulate the first metric to quantify the enterprise network exposure in this context, namely the Enterprise Network Internal Connectivity Exposure (ENICE). We compute the ENICE metric following Equation \ref{eq:ENICE} where $w(a)$ is the weight of edge $a \in A$.
\begin{equation}\label{eq:ENICE}
 ENICE = \sum\limits_{a\in A} w(a)   
\end{equation}

The clustering coefficient of a connectivity graph measures how interconnected a vertex’s neighbours are to one another. The higher the clustering coefficient, the more exposed the network assets are to lateral attacker movement. The global clustering coefficient is designed to give an overall indication of the clustering in the network based on triplets of nodes. A triplet is three nodes that are connected by either two (open triplet) or three (closed triplet) ties. Accordingly, the second exposure metric, namely the Global Clustering coefficient (GC), is calculated as follows:

\begin{equation}\label{eq:GCC}
    GC_C={\frac{number\;of\;closed\;triplets}{total\;number\;of\; triplets}}
\end{equation}

\noindent\textbf{Reachability: }Given the network connectivity graph, finding out the vertices that are reachable from another vertex $\mathit{v}$ is an indicator of the number of possibly compromised assets in case $\mathit{v}$ is illegitimately accessed. A direct measure of reachability is the shortest path analysis of the network connectivity graph. The shortest path represents the minimum number of hops an attacker needs to exploit until reaching a target asset. The longer the shortest path, the more effort an attacker needs to deploy to compromise a target network asset. The Mean of shortest Path Length (MPL) represents the average number of hosts, in the best case, an attacker needs to compromise in order to reach their target. It is a strong indicator of how likely are the network assets reachable from a compromised source. Let us denote by $\mathit{LSP_C}$ the list of shortest paths in the connectivity graph $\mathit{C}$ and by $\mathit{p}$ a path in the connectivity graph where $\mathit{|p|}$ is the number of vertices in the path $\mathit{p}$. $\mathit{MLP_C}$ is calculated following Equation \ref{eq:avgp}.
\begin{equation}\label{eq:avgp}
    MPL_C={\frac{1}{|LSP_C|}\sum\limits_{\forall p \in LSP_C}|p|}
\end{equation}

The diameter of a graph is the maximum eccentricity of any vertex in the graph. It is equal to the maximum shortest path length of the connectivity graph. It indicates the longest chain an attacker is forced to pursue to compromise a target network asset from a compromised source. The latter assumes that attackers usually aim at taking the shortest path to reach their target asset. Let $s(u,v)$ be the shortest path distance from vertex $u$ to vertex $v$. The Connectivity graph Diameter (CD), the second reachability metric, can be defined by:

\begin{equation}\label{eq:MITCO}
    CD_C=\max_{\forall v,u \in V}{\left \{ s(u, v) \right\}}
\end{equation}

The transitive closure of a connectivity graph $\mathit{C}$ is a graph $\mathit{C^T=(A^T,V)}$ such that for all $\mathit{i,j \in V}$ there is a link $\mathit{(i,j)}$ if and only if there exists a path from $\mathit{i}$ to $\mathit{j}$ in $\mathit{C}$. Therefore, it provides a more detailed perspective of how far an attacker can go after compromising a network asset. A direct indicator of the latter is the number of transitive paths in the network represented by the transitive closure graph edges. Standing on this ground, we proceed by formulating the third reachability metric, namely the Transitive Internal Network Reachability (TINR), following Equation \ref{eq:TINE} where $\mathit{A^T}$ is the set of edges of the transitive closure graph $\mathit{C^T}$ corresponding to the network connectivity graph $\mathit{C}$.
\begin{equation}\label{eq:TINE}
    TINR_C =|A^T|
\end{equation}

\noindent\textbf{Centrality: }Nodal centrality quantifies how important a node is within a network. Hence, it implies the impact of compromising the node on the overall network security. The out-degree centrality metric of the network connectivity graph vertex $\mathit{v}$ outlines the number of possibly compromised assets by the attacker in case $\mathit{v}$ is compromised. Accordingly, we develop a metric namely AVerage Out-Degree (AVOD) to account for the typical number of possibly compromised assets after a successful attack. Let $\mathit{OD(v)}$ be the out-degree of a connectivity graph $\mathit{C(A,V)}$ node $\mathit{v \in V}$. The average out-degree is given by:
\begin{equation}\label{eq:avgOD}
    AVOD_C={\frac{1}{|V|}\sum\limits_{\forall v \in V}OD(v)}
\end{equation}

The closeness centrality is tightly related to the notion of distance between nodes. It highlights nodes that may reach any other nodes within a few hops and nodes that may be very distant. The closeness centrality of a network asset $\mathit{v}$ indicates how fast an attacker, after compromising $\mathit{v}$, can access all other nodes in the network. The CLoseness centrality (CL) of node $\mathit{v}$ is computed following Equation \ref{eq:cl} where $\mathit{d(v, u)}$ is the distance (number of vertices) between $v$ and $u$:
\begin{equation}\label{eq:cl}
    CL(v)={\frac{1}{\sum\limits_{\forall u \in V}d(v,u)}}
\end{equation}
The average closeness of the network connectivity graph nodes, the second centrality metric, is an indicator of how fast the whole network can be compromised after a breach. It is calculated as follows:
\begin{equation}\label{eq:avcl}
    AC_C={\frac{1}{|V|}\sum\limits_{\forall v \in V}CL(v)}
\end{equation}

\subsection{Network Robustness}\label{sec:Risk} 

Network robustness is the ability of a network to resist goal-oriented attackers. To measure the network robustness, we leverage the attack graph structure in conjunction with a component metric such as the Common Vulnerability Scoring System (CVSS) \cite{cvss1,cvss2}. An attack graph is a succinct representation of all paths through a system that ends in a state where an intruder has successfully achieved his goal \cite{attack}. We classify the attack graph-based network robustness security metrics into three categories; non-path-based, path-based and CVSS-based security metrics. We generate the attack graph using the MulVAL tool \cite{mulval1, mulval2, mulval3} while relying on the Nessus \cite{nessus} vulnerabilities scanner output and the enterprise network perimeter and micro-segmentation firewall rules.

%\noindent\textbf{Problem formulation: } 
To formally define network robustness, let $\mathit{G(E, N)}$ be the enterprise network attack graph consisting of a set of nodes $\mathit{N}$ of three types: attack step nodes, privilege nodes and configuration nodes \cite{mulval4}. Let $\mathit{R_G}$ be the root nodes of the attack graph $\mathit{G}$ representing network configurations that contribute to attack possibilities. Let $\mathit{L_G}$ be the set of privilege nodes denoting the compromised assets. The set of paths $\mathit{P_G}$ of the attack graph $\mathit{G}$ comprises all directed attack paths starting at the root configuration nodes $\mathit{R_G}$ and ending at the privilege nodes $\mathit{L_G}$. 

\noindent\textbf{Path-Based metrics: }The shortest path metric indicates the minimum number of attack steps an attacker should perform to compromise an asset in the network. Since a chain is only as strong as its weakest link, it is a significant indicator of the network robustness. Indeed, attackers would need a global view of the system vulnerabilities to deliberately exploit the shortest path. However, the probability of the shortest path to be exploited is directly proportional to the number of shortest paths in the network \cite{attack6}. Hence, we define the Number of Shortest Paths metric (NSP) to identify the count of shortest attack paths between every root node $\mathit{r \in R_G}$ and privilege node $\mathit{l \in L_G}$ in the attack graph $\mathit{G}$. Let us denote by $\mathit{LSAP_G}$ the list of shortest attack paths in the attack graph $\mathit{G}$. The $\mathit{NSP}$ metric is given by:
\begin{equation}\label{eq:nshortp}
    NSP_G=|LSAP_G|
\end{equation}

The Minimum Shortest Path Length metric (MSPL) denotes the absolute minimum number of attack steps an attacker needs to perform to compromise a target network asset. Let $\mathit{p}$ be a path in the attack graph where $\mathit{|p|}$ is the number of nodes in $\mathit{p}$, accordingly: 

\begin{equation}\label{eq:minsp}
    MSPL_G=\min_{\forall p \in LSAP_G}{\left \{ |p| \right\}}
\end{equation}

The count of paths with length equal to the minimum shortest path length outlines the number of network weakest links. Hence, in conjunction with the MSPL metric, it is a direct indicator of the likelihood of a successful attack. Consequently, we define the Count of Minimum Path Length (CMPL) metric. Let $\mathit{LMP_G}$ be the list of paths with length equals to $\mathit{MSPL_G}$ in the attack graph $\mathit{G}$. Accordingly, $\mathit{CMPL}$ is given by:
\begin{equation}\label{eq:Nminp}
    CMPL_G=|LMP_G|
\end{equation}

%-----------------------------------------------------------------
\noindent\textbf{Non-Path-Based metrics: }The count of the attack graph configuration nodes is an intuitive yet significant non-path-based measure of the network robustness. The configuration nodes depict facts about the current network configuration that contributes to one or more attack possibilities. Hence, they stand as the attacker's entry point to the network. We proceed by formulating the first non-path metric namely the Count of Mis-Configurations (CMC) following Equation \ref{eq:CMC} where $\mathit{R_G}$ is the set of root nodes of the attack graph $\mathit{G}$:
\begin{equation}\label{eq:CMC}
    CMC_G=|R_G|
\end{equation}

The out-degree metric (OD) of the attack graph privilege nodes $\mathit{L_G}$ is an indicator of the number of possible attacks succeeding a compromised network asset. Hence, we define the average out-degree $\mathit{AOD_G}$ and the maximum out-degree $\mathit{MOD_G}$ metrics. While the earlier is an indicator of the typical number of attacks after a compromised asset the latter represents the worst-case scenario, indicating the maximum number of possible attacks. 

Let $\mathit{OD(n)}$ be the out-degree of an attack graph privilege node $\mathit{n}$. The Average Out-Degree (AOD) and the Maximum Out-Degree (MOD) metrics of an attack graph $\mathit{G}$ are given by Equations \ref{eq:avgODA} and \ref{eq:maxOD}, respectively.

\begin{equation}\label{eq:avgODA}
    AOD_G={\frac{1}{|L_G|}\sum\limits_{\forall n \in L_G}OD(n)}
\end{equation}

\begin{equation}\label{eq:maxOD}
    MOD_G=\max_{\forall n \in L_G}{\left \{ OD(n) \right\}}
\end{equation}

The betweenness centrality of an attack privilege node is the extent to which the vertex plays a bridging role in a network. In other words, it measures the extent that the privilege node falls on the shortest path to other privilege nodes. Accordingly, the higher the betweenness of a compromised privilege node, the more widespread across the network the subsequent privileges the attacker can acquire. The betweenness metric further provides insight into the network weak links that should be prioritised for redemption. The betweenness centrality of a privilege node $\mathit{n}$ is calculated following Equation \ref{eq:btn} where $\mathit{NSP_{rl}}$ is the number of shortest paths from root $\mathit{r \in R_G}$ to privilege $\mathit{l \in L_G}$, $\mathit{NSP_{rl}(n)}$ is the number of shortest paths passing through $\mathit{n \in L_G}$ and $\mathit{r \neq l \neq n}$. 

\begin{equation}\label{eq:btn}
    BN(n)={\sum\limits_{ \forall r \in R_G \land  \forall l \in L_G}\frac{ NSP_{rl}(n)}{ NSP_{rl}}}
\end{equation}

The Average Betweenness (AB) of the attack graph privilege nodes depicts the typical contribution of an illegitimately acquired network privilege in other potential attacks across the network compromising privileges that the attacker could not have reached otherwise. %In other words, its propensity of being an attack step in more sophisticated attacks widespread across the network. 
It is calculated following Equation \ref{eq:Avfbtn} where $\mathit{L_G}$ is the set of privilege nodes of the attack graph $\mathit{G}$. 
\begin{equation}\label{eq:Avfbtn}
    AB_G={\frac{1}{|L_G|}\sum\limits_{\forall l \in L_G}BN(l)}
\end{equation}

\noindent\textbf{Common Vulnerability Scoring System (CVSS) metrics: }Assigning complexity values of the network vulnerabilities communicates their characteristics and severity. This approach reflects variations in the difficulty of exploiting the different vulnerabilities \cite{Wang}. A standard, such as the CVSS \cite{cvss1}, may be used to provide guidance in scoring vulnerabilities. 

While there is currently no standard way of aggregating vulnerability metrics, a critical issue in measuring network security is to combine measures of individual vulnerabilities, and configurations into a global measure \cite{aggregate3}. We rely on the work presented in \cite{aggregate2, aggregate, aggregate3} to analyse the enterprise network attack graph for the purpose of calculating cumulative metrics and aggregating the vulnerabilities score. 

The cumulative score of a given privilege node indicates the likelihood that the corresponding resource is compromised during an attack, or equivalently, among all attackers attacking the network over a given time period, the average fraction of attackers who can successfully compromise the resource, taking into consideration the effects of all possible inter-plays between vulnerabilities. We refer the reader to \cite{aggregate} for further details on the underlying mathematical modelling and computation of risk.

\section{Analysis and Results}\label{sec:aresults}
In this section, we assess the effectiveness of the proposed framework in evaluating the impact of micro-segmentation on enterprise network security. To construct the network connectivity graph, representing the basis of the exposure analysis, we identify the network hosts forming the graph nodes. While the flat network connectivity graph is a complete graph, we rely on the micro-segmentation firewall rules to identify the edges of the segmented network connectivity graph. The firewall rules are described in terms of the source host, destination host, service protocol and service destination port.

To assess the network robustness, the attack graph is generated using the MulVAL tool.  It is given as input the Nessus XML output and the network firewall rules. While the earlier depicts the hosts' vulnerabilities and configurations including the running software and services, the latter models the assets access control policies regulating their communication. The firewall rules include the micro-segmentation rules, if any, in addition to the enterprise network firewall rules regulating north-south traffic. In the case of a flat network, the micro-segmentation rules are replaced by one generic rule allowing all internal network traffic.

We consider the network data of two enterprises; a university and a life-care organisation. Table \ref{tab:stats} provides statistics describing the two networks. The connections depict the count of asset pairs that are allowed to communicate. While we were able to assess the exposure of both enterprises, due to the unavailability of the Nessus data for the life-care organisation (Enterprise B), the robustness analysis was only limited to the university enterprise network (Enterprise A).

The metrics are calculated prior to and post the deployment of micro-segmentation. The results are then compared, assessed and used to draw conclusions which lead to useful insights that follow in this section.

\begin{table*}[ht]
\centering
%\caption{Network configuration}
\caption{Summary of the dataset consisting of two networks' configurations}
\label{tab:stats}
%\resizebox{\linewidth}{!}{

\begin{tabular}{|c||c|c|c|c|}
\hline
%\textbf{Enterprise} & \textbf{Business} & \textbf{Hosts} & \textbf{Flat Connections} & \textbf{Segmented Connections} \\
\textbf{Enterprise} & \textbf{Business} & \textbf{\# of Hosts} & \textbf{\# Connections (Flat Network)} & \textbf{\# of Connections (Segmented Network)} \\
\hline

\hline
Enterprise A & University & 300 & 90,000 & 4,045 \\
Enterprise B  & Life-care organization & 238 & 56,644 & 3,007  \\

\hline
\end{tabular}
%}
\vspace{-0.4cm}
\end{table*}

\subsection{Network Exposure Analysis}
\label{subsec:expoanalysis}
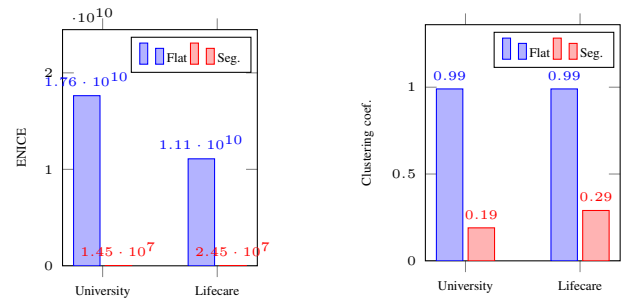
\begin{figure}[!ht]
\vspace{-0.3cm}
\subfloat[Network Internal Exposure]{
	\begin{minipage}[c][1\width]{
	   0.23\textwidth}
	   \centering
	   	   \begin{tikzpicture}
            \begin{axis}[  
    ybar,  
    enlarge y limits={0.36,upper},
    enlarge x limits=0.35,  
    legend pos=north east,
    legend columns=2,
    legend style={nodes={scale=0.5, transform shape}},
    ymax=18000000000,
    ymin=0,
    ylabel={\tiny ENICE}, % there should be no line gap between the rows here. Otherwise, latex will show an error.  
    yticklabel style={font=\tiny },
    xticklabel style={font=\tiny },
    ytick style={font=\tiny },
    symbolic x coords={University, Lifecare},  
    xtick=data,  
    nodes near coords,  
    nodes near coords align={vertical}, 
    nodes near coords style={font=\tiny },
    width=\textwidth,height=0.2\textheight
    ]  
            \addplot+ coordinates {(University, 17635468500) (Lifecare, 11089701630)}; 
            \addplot+ coordinates {(University, 14523704) (Lifecare, 24489385)}; 
            
            \legend{Flat, Seg.}; \end{axis} 
        \end{tikzpicture} 
	\end{minipage}\label{fig:enice}}
 \hfill 	
  \subfloat[Global Clustering Coef.]{
	\begin{minipage}[c][1\width]{
	   0.23\textwidth}
	   \centering
	   \begin{tikzpicture}
            \begin{axis}[  
    ybar, % ybar command displays the graph in horizontal form, while the xbar command displays the graph in vertical form.  
    enlarge y limits={0.36,upper},
    enlarge x limits=0.35,
    legend pos=north east,
    legend columns=2,
    legend style={nodes={scale=0.5, transform shape}},
    ymax=1,
    ymin=0,  
    ylabel={\tiny Clustering coef.}, % there should be no line gap between the rows here. Otherwise, latex will show an error.  
    yticklabel style={font=\tiny },
    xticklabel style={font=\tiny },
    ytick style={font=\tiny },
    symbolic x coords={University, Lifecare},  
    xtick=data,  
    nodes near coords,  
    nodes near coords align={vertical}, 
    nodes near coords style={font=\tiny },
    width=\textwidth,height=0.2\textheight
    ]  
            \addplot+ coordinates {(University, 0.99) (Lifecare, 0.99)}; 
            \addplot+ coordinates {(University, 0.19) (Lifecare, 0.29)}; 
            
            \legend{Flat, Seg.}; \end{axis} 
        \end{tikzpicture} 
	\end{minipage}\label{fig:clstc}}
\caption{Connectedness metrics analysis.}
%\caption{Connectivity and risk exposure analysis.}
%\vspace{-0.3cm}
\label{fig:connectedness}
\end{figure}

%After our developed framework constructs the connectivity graphs of the flat network and segmented network, respectively, it automatically runs the exposure metrics module. 
%The exposure metrics are calculated twice; once prior to the deployment of micro-segmentation and again after the deployment. The results are then compared, assessed and used to draw conclusions which lead to useful insights that follow in this section.\\
%----------------------------------------connectedness
\noindent\textbf{Connectedness: }After micro-segmentation deployment, the ENICE values of Enterprises A and B networks are decremented by 99.92\% and 99.78\% respectively as shown in Figure \ref{fig:enice}. Since the flat network topology, by definition, allows all internal network communications, we assumed that devices can listen on any port. Hence, for the flat network topology, we considered that all connections have an equivalent weight of 65,535. It can be argued that hosts are not necessarily listening on every possible port. However, in the flat network topology, nothing can stop a network asset from listening on a particular port. In fact, not many organizations, prior to deploying micro-segmentation, are fully aware of all the ports their assets are listening on. Therefore, micro-segmentation not only regulates the hosts that are allowed to directly communicate, it further restricts the services used for communication between the different network assets. 

Restricting the network connectedness further minimises the impact of the zero-day attacks by limiting the ports the micro-segments are listening on. Let us consider the zero-day threat caused by the Windows vulnerability SMBGhost \cite{SMBGOHST} affecting the Microsoft Server Message Block (SMB) of the network file sharing protocol. Successful exploitation of this vulnerability enables the attacker to execute arbitrary code on the SMB server or client. The attacker could then possibly install programs, view, edit or remove data. Deploying micros-segmentation, the control rules deny the file-sharing capability on all the enterprise database servers while only allowing connectivity on the SQL port. Despite the presence of a zero-day vulnerability that could be exploited, no attacker is able to connect to the database server on the port that is vulnerable.

Similar behaviour is depicted by the global clustering coefficient. As shown in Figure \ref{fig:clstc}, for the flat networks, the GC metric values are close to one indicating a high percentage of closed triads in the graphs. The latter implies that the graph nodes are involved in as many transitive relations as possible. On the other hand, the micro-segmented network GC values are reduced by 80\% and 70\% for Enterprise A and Enterprise B, respectively. It can be remarked that Enterprise A clustering coefficient improvement percentage is higher than B. This can be attributed to the fact that B has leveraged infrastructure across applications thus allowing more transitive connections. In other words, Enterprise B has fewer dedicated unique servers per application. For example, it has three applications sharing the same database server.

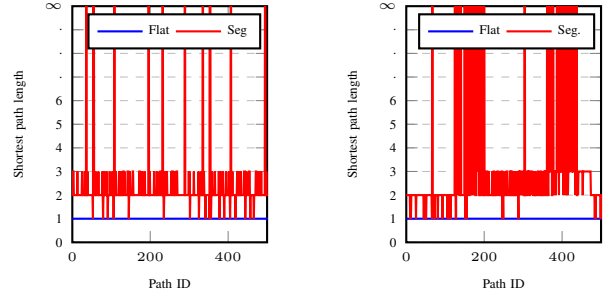
\begin{figure}[ht]
  \subfloat[University Network]{
	\begin{minipage}[c][1\width]{
	   0.23\textwidth}
	   \centering
	   \begin{tikzpicture}
            \tikzset{
            every pin/.style={fill=yellow!50!white,rectangle,rounded corners=3pt,font=\tiny},
            small dot/.style={fill=black,circle,scale=0.3}
            }
            
            \begin{axis}[
            width=1\textwidth,
            height=0.2\textheight,
            ymajorgrids=true,
                grid style=dashed,
                xlabel={\tiny Path ID},
                ylabel={\tiny Shortest path length},
            xmin=0, 
            xmax=500,
            ymax=10,
            ymin=0,
            ytick distance=1,
            %ymode=log,
            %xtick distance=1,
            legend pos=north east,
            legend columns=2,
            legend style={nodes={scale=1, transform shape}},
            yticklabel style={font=\tiny },
            xticklabel style={font=\tiny },
            yticklabels={0, 0, 1, 2, 3, 4, 5, 6, ., ., .,$\infty$},
            no marks,
            thick
            ]
            \addplot table [col sep=comma,trim cells=true,x=ID,y=flat] {graphs/pathlength3.csv};
            \addplot table [col sep=comma,trim cells=true,x=ID,y=Seg] {graphs/pathlength3.csv};
            
            %\node[small dot,pin=200:{$(5,0.12)$}] at (axis cs:5,0.12) {};
            
            \legend{\tiny Flat, \tiny Seg }
            %%\addplot[mark=*] coordinates {(5,0.12)} node[pin=150:{$(5,0.12)$}]{}
            \end{axis}
        \end{tikzpicture}
	\end{minipage}
	\label{fig:shortestPathFlat}}
 \hfill 	
  \subfloat[Life-care Network]{
	\begin{minipage}[c][1\width]{
	   0.23\textwidth}
	   \centering
	   	   \begin{tikzpicture}
                \tikzset{
                every pin/.style={fill=yellow!50!white,rectangle,rounded corners=3pt,font=\tiny},
                small dot/.style={fill=black,circle,scale=0.3}
                }
                
                \begin{axis}[
                width=1\textwidth,
                height=0.2\textheight,
                ymajorgrids=true,
                    grid style=dashed,
                    xlabel={\tiny Path ID},
                    ylabel={\tiny Shortest path length},
                xmin=0, 
                xmax=500,
                ymax=10,
                ymin=0,
                ytick distance=1,
                %ymode=log,
                %xtick distance=1,
                legend pos=north east,
                legend style={nodes={scale=1, transform shape}},
                legend columns=2,
                yticklabel style={font=\tiny },
                xticklabel style={font=\tiny },
                yticklabels={0, 0, 1, 2, 3, 4, 5, 6, ., ., .,$\infty$},
                no marks,
                thick
                ]
                \addplot table [col sep=comma,trim cells=true,x=ID,y=flat] {graphs/pathlength_rsl.csv};
                \addplot table [col sep=comma,trim cells=true,x=ID,y=Seg] {graphs/pathlength_rsl.csv};
                
                %\node[small dot,pin=200:{$(5,0.12)$}] at (axis cs:5,0.12) {};
                
                \legend{\tiny Flat, \tiny Seg. }
                %%\addplot[mark=*] coordinates {(5,0.12)} node[pin=150:{$(5,0.12)$}]{}
                \end{axis}
            \end{tikzpicture}
	\end{minipage}
	\label{fig:shortestPathSeg}}
\caption{Connectivity graph shortest path length distributions.}% \ik{Increase font size of legends}}
%\caption{Shortest path length (i.e., in number of hops shown in y-axis) of the connectivity graphs for both Flat and Segmented Networks. }% \ik{Increase font size of legends}}
\vspace{-0.3cm}

\label{fig:shortestPath}
\end{figure}

%_____________________________reachability
\noindent\textbf{Reachability}: We begin by analysing the distribution of the shortest paths of the network connectivity graphs. Figure \ref{fig:shortestPath} presents the paths length distribution before and after micro-segmentation. The x-axis denotes the relation identifier of the possibly connected pair of hosts and the y-axis represents the shortest path length connecting the two nodes, if any. The shortest path length of the flat network of both enterprises has a constant value of one. Hence, an attacker can reach any other network asset in one step. This represents the situation of an employee laptop directly communicating with an enterprise database. In the event the laptop is infected with malware or CryptoLocker, it could directly infect the connected databases. Contrastingly, the majority of paths length in the segmented network of both enterprises falls in the range of $[2,3]$ with an average value of 2.14 for Enterprise A and 2.17 for Enterprise B. It should be noted that the shortest path length values reflect the number of IT services tiers the organization have. For the two hereby considered enterprises, after micro-segmentation, services have either three tiers of workloads (user $\rightarrow$ web server, web server $\rightarrow$ application server, application server $\rightarrow$ database) or two tiers (when the web server and application server tiers are combined). Consequently, the typical effort an attacker needs to deploy in order to reach a target asset is doubled as a result of micro-segmentation deployment. 

The paths with length going to infinity characterising the segmented network of both enterprises designate that there is no path between the two possibly connected assets. The percentage of infinity paths of the segmented Enterprises A and B networks are 5.6\% and 27.3\% respectively which further limits the assets reachability by restricting the attacker's lateral movement within the network. It is worth mentioning that Enterprise A has fewer infinity paths because it has more management services characterised by their high connectivity.

Similar behaviour is observed with the worst-case reachability value depicted by the Connectivity graph Diameter metric (CD). Since the connectivity graph of the flat network is complete, the attacker can reach any target asset in exactly one step. Contrastingly, in the case of the micro-segmented network, the maximum effort deployed to attain the intended asset is tripled for both enterprises. This can be explained by the fact that, for these two enterprises, IT services are divided into a maximum of three tiers. Therefore, to compromise the database tier, the attacker needs to first gain access to the web server and the application server tiers.

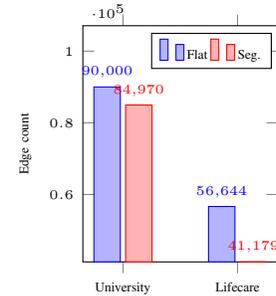
\begin{wrapfigure}{R}{0.23\textwidth}
	\begin{minipage}[c][1\width]{
	   0.23\textwidth}
	   \centering
	   	   \begin{tikzpicture}
            \begin{axis}[  
    ybar, % ybar command displays the graph in horizontal form, while the xbar command displays the graph in vertical form.  
    enlarge y limits={0.35,upper},
    enlarge x limits=0.35,
    legend pos=north east,
    legend columns=2,
    legend style={nodes={scale=0.5, transform shape}},
    ymax=90000,
    ylabel={\tiny Edge count}, % there should be no line gap between the rows here. Otherwise, latex will show an error.  
    yticklabel style={font=\tiny },
    xticklabel style={font=\tiny },
    ytick style={font=\tiny },
    symbolic x coords={University, Lifecare},  
    xtick=data,  
    nodes near coords,  
    nodes near coords align={vertical}, 
    nodes near coords style={font=\tiny },
    width=\textwidth,height=0.2\textheight
    ]  
            \addplot+ coordinates {(University, 90000) (Lifecare, 56644)}; 
            \addplot+ coordinates {(University, 84970) (Lifecare, 41179)}; 
            
            \legend{Flat, Seg.}; \end{axis} 
        \end{tikzpicture} 
	\end{minipage}
\caption{Transitive closure edges count.}
\label{fig:transitive}
\vspace{-0.3cm}
\end{wrapfigure}

Whilst the impact of micro-segmentation on the mean shortest path length and diameter of both enterprises is almost identical, the TINR metric exhibits a different behaviour as highlighted by Figure \ref{fig:transitive}. For Enterprise A, the transitive reachability is decreased by 5.6\% only whereas for Enterprise B, it is decreased by 27.3\%. The latter values are found to be identical to the percentages of disconnected nodes as a result of deploying micro-segmentation. Hence, it can be safely deduced that the risk of compromising the network assets is a factor of the number of allowed internal network connections.

%------------------- paths length distribution

%-------------------reachability characherization

%-------------------centrality characterization

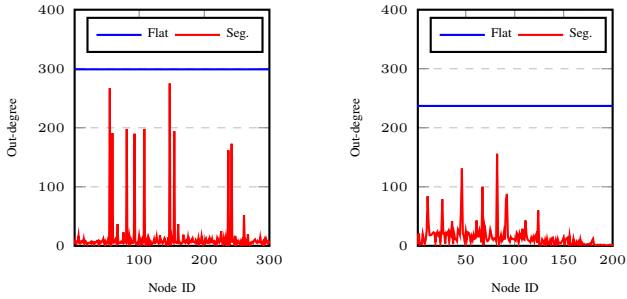
\begin{figure}[ht]
  \subfloat[University Network]{
	\begin{minipage}[c][1\width]{
	   0.23\textwidth}
	   \centering
	   \begin{tikzpicture}
            \tikzset{
            every pin/.style={fill=yellow!50!white,rectangle,rounded corners=3pt,font=\tiny},
            small dot/.style={fill=black,circle,scale=0.3}
            }
            
            \begin{axis}[
            width=1\textwidth,
            height=0.2\textheight,
            ymajorgrids=true,
                grid style=dashed,
                xlabel={\tiny Node ID},
                ylabel={\tiny Out-degree},
            xmin=1, 
            xmax=300,
            ymax=400,
            ymin=0,
            %ymode=log,
            %xtick distance=1,
            legend pos=north east,
            legend columns=2,
            legend style={nodes={scale=1, transform shape}},
            yticklabel style={font=\tiny },
            xticklabel style={font=\tiny },
            no marks,
            thick
            ]
            \addplot table [col sep=comma,trim cells=true,x=ID,y=outDegflat] {graphs/outDegrees.csv};
            \addplot table [col sep=comma,trim cells=true,x=ID,y=outDegSeg] {graphs/outDegrees.csv};
            \legend{\tiny Flat , \tiny Seg.  }
            %%\addplot[mark=*] coordinates {(5,0.12)} node[pin=150:{$(5,0.12)$}]{}
            \end{axis}
        \end{tikzpicture}
	\end{minipage}}
 \hfill 	
  \subfloat[Life-care Network]{
	\begin{minipage}[c][1\width]{
	   0.23\textwidth}
	   \centering
	   \begin{tikzpicture}
            \tikzset{
            every pin/.style={fill=yellow!50!white,rectangle,rounded corners=3pt,font=\tiny},
            small dot/.style={fill=black,circle,scale=0.3}
            }
            
            \begin{axis}[
            width=1\textwidth,
            height=0.2\textheight,
            ymajorgrids=true,
                grid style=dashed,
                xlabel={\tiny Node ID},
                ylabel={\tiny Out-degree},
            xmin=1, 
            xmax=200,
            ymax=400,
            ymin=0,
            %ymode=log,
            %xtick distance=1,
            legend pos=north east,
            legend columns=2,
            legend style={nodes={scale=1, transform shape}},
            yticklabel style={font=\tiny },
            xticklabel style={font=\tiny },
            no marks,
            thick
            ]
            \addplot table [col sep=comma,trim cells=true,x=ID,y=outDegflat] {graphs/outDegrees_rsl.csv};
            \addplot table [col sep=comma,trim cells=true,x=ID,y=outDegSeg] {graphs/outDegrees_rsl.csv};
            \legend{\tiny Flat, \tiny Seg. }
            %%\addplot[mark=*] coordinates {(5,0.12)} node[pin=150:{$(5,0.12)$}]{}
            \end{axis}
        \end{tikzpicture}
	\end{minipage}}
	
\caption{Connectivity graph out-degree distribution.}
%\caption{Connectivity graph out-degree distribution of Enterprises A \& B).}
\label{fig:outdegree}
\vspace{-0.4cm}
\end{figure}

\noindent\textbf{Centrality: }The flat topology, by definition, allows all internal network traffic resulting in a complete connectivity graph. If any node is compromised the whole network is at risk as depicted by the linear out-degree distribution of Figure \ref{fig:outdegree}. In contrast, micro-segmentation significantly reduces the average out-degree centrality by 95.5\% and 94.7\% for Enterprise A and Enterprise B, respectively, as shown in Figure \ref{fig:avgOD}.

Despite the undeniable improvement of micro-segmentation on the overall network out-degree centrality, Figure \ref{fig:outdegree} exhibits unexpected behaviour of the values of the individual nodes. It can be noticed that some nodes have a high out-degree and can be a source of risk if compromised. After further analysis, they are found to belong to common management services (e.g. active directory, backup, ntp, dns, etc.) characterised by high connectivity. It should be noted that the hereby presented framework highlights, in different contexts, the network weak links that need to be given priority for investigation and mitigation.

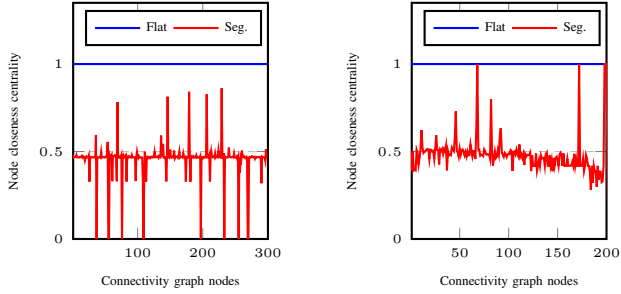
\begin{figure}[ht]
  \subfloat[University Network]{
	\begin{minipage}[c][1\width]{
	   0.23\textwidth}
	   \centering
	   \begin{tikzpicture}
            \tikzset{
            every pin/.style={fill=yellow!50!white,rectangle,rounded corners=3pt,font=\tiny},
            small dot/.style={fill=black,circle,scale=0.3}
            }
            
            \begin{axis}[
            enlarge y limits={0.35,upper},
            width=1\textwidth,
            height=0.2\textheight,
            ymajorgrids=true,
                grid style=dashed,
                xlabel={\tiny Connectivity graph nodes},
                ylabel={\tiny Node closeness centrality},
            xmin=1, 
            xmax=300,
            ymax=1,
            ymin=0,
            %ymode=log,
            %xtick distance=1,
            legend pos=north east,
            legend columns=2,
            legend style={nodes={scale=1, transform shape}},
            yticklabel style={font=\tiny },
            xticklabel style={font=\tiny },
            no marks,
            thick
            ]
            \addplot table [col sep=comma,trim cells=true,x=ID,y=flat] {graphs/closeness.csv};
            \addplot table [col sep=comma,trim cells=true,x=ID,y=Seg] {graphs/closeness.csv};
            
            %\node[small dot,pin=200:{$(5,0.12)$}] at (axis cs:5,0.12) {};
            
            \legend{\tiny Flat , \tiny Seg.  }
            %%\addplot[mark=*] coordinates {(5,0.12)} node[pin=150:{$(5,0.12)$}]{}
            \end{axis}
        \end{tikzpicture}
	\end{minipage}}
 \hfill 	
  \subfloat[Life-care Network]{
	\begin{minipage}[c][1\width]{
	   0.23\textwidth}
	   \centering
	   \begin{tikzpicture}
            \tikzset{
            every pin/.style={fill=yellow!50!white,rectangle,rounded corners=3pt,font=\tiny},
            small dot/.style={fill=black,circle,scale=0.3}
            }
            
            \begin{axis}[
           enlarge y limits={0.35,upper},
            width=1\textwidth,
            height=0.2\textheight,
            ymajorgrids=true,
                grid style=dashed,
                xlabel={\tiny Connectivity graph nodes},
                ylabel={\tiny Node closeness centrality},
            xmin=1, 
            xmax=200,
            ymax=1,
            ymin=0,
            %ymode=log,
            %xtick distance=1,
            legend pos=north east,
            legend columns=2,
            legend style={nodes={scale=1.0, transform shape}},
            yticklabel style={font=\tiny },
            xticklabel style={font=\tiny },
            no marks,
            thick
            ]
            \addplot table [col sep=comma,trim cells=true,x=ID,y=flat] {graphs/closeness_rsl.csv};
            \addplot table [col sep=comma,trim cells=true,x=ID,y=Seg] {graphs/closeness_rsl.csv};
            
            %\node[small dot,pin=200:{$(5,0.12)$}] at (axis cs:5,0.12) {};
            
            \legend{\tiny Flat, \tiny Seg. }
            %%\addplot[mark=*] coordinates {(5,0.12)} node[pin=150:{$(5,0.12)$}]{}
            \end{axis}
        \end{tikzpicture}
	\end{minipage}
	}
\caption{Closeness centrality distribution.}% \ik{increase legends' font size}}
%\caption{Nodes' closeness centrality analysis of the Flat and Segmented networks connectivity graphs of the two analysed enterprises.}
\vspace{-0.4cm}
\label{fig:closeness}
\end{figure}

The closeness centrality distribution values presented in Figure \ref{fig:closeness} fall in the range of [0,1]. A value close to zero indicates that a given node is distant from other nodes in the network. It further signifies that numerous links need to be traversed to get to other nodes in the network. It can be remarked that the flat closeness value for both enterprises is equal to one. This can be attributed to the fact that all flat network nodes are exactly one hop away distant from each other. Therefore, after acquiring an illegitimate privilege, an attacker can access any other asset by traversing only one link. On the other hand, after micro-segmentation, the overall closeness is reduced by almost half for both enterprises. The high spikes characterising the individual nodes' closeness distribution of Figure \ref{fig:closeness} belong to common management services. 

%------------------- outdegree figure

%------------------- closeness figure

%-------------------centrality characherization
\begin{figure}[!ht]
  \subfloat[Average Out-Degree]{
	\begin{minipage}[c][1\width]{
	   0.23\textwidth}
	   \centering
	   \begin{tikzpicture}
            \begin{axis}[  
    ybar, % ybar command displays the graph in horizontal form, while the xbar command displays the graph in vertical form.  
    enlarge y limits={0.36,upper},
    enlarge x limits=0.35,
    legend pos=north east,
    legend columns=2,
    legend style={nodes={scale=0.5, transform shape}},
    ymax=300,
    ymin=0,  
    ylabel={\tiny Average out-degree}, % there should be no line gap between the rows here. Otherwise, latex will show an error.  
    yticklabel style={font=\tiny },
    xticklabel style={font=\tiny },
    ytick style={font=\tiny },
    symbolic x coords={University, Lifecare},  
    xtick=data,  
    nodes near coords,  
    nodes near coords align={vertical}, 
    nodes near coords style={font=\tiny },
    width=\textwidth,height=0.2\textheight
    ]  
            \addplot+ coordinates {(University, 299) (Lifecare, 237)}; 
            \addplot+ coordinates {(University, 13) (Lifecare, 12)}; 
            
            \legend{Flat, Seg.}; \end{axis} 
        \end{tikzpicture} 
	\end{minipage}
	\label{fig:avgOD}}
 \hfill 	
  \subfloat[Average Closeness]{
	\begin{minipage}[c][1\width]{
	   0.23\textwidth}
	   \centering
	   	   \begin{tikzpicture}
            \begin{axis}[  
    ybar, % ybar command displays the graph in horizontal form, while the xbar command displays the graph in vertical form.  
    enlarge y limits={0.36,upper},
    enlarge x limits=0.35,  
    legend pos=north east,
    legend columns=2,
    legend style={nodes={scale=0.5, transform shape}},
    ymax=1,
    ymin=0,
    ylabel={\tiny Average closeness}, % there should be no line gap between the rows here. Otherwise, latex will show an error.  
    yticklabel style={font=\tiny },
    xticklabel style={font=\tiny },
    ytick style={font=\tiny },
    symbolic x coords={University, Lifecare},  
    xtick=data,  
    nodes near coords,  
    nodes near coords align={vertical}, 
    nodes near coords style={font=\tiny },
    width=\textwidth,height=0.2\textheight
    ]  
            \addplot+ coordinates {(University, 1) (Lifecare, 1)}; 
            \addplot+ coordinates {(University, 0.45) (Lifecare, 0.4)}; 
            
            \legend{Flat, Seg.}; \end{axis} 
        \end{tikzpicture} 
	\end{minipage}
	\label{fig:Avgcl}}
\caption{Connectivity graph Centrality analysis.}
%\caption{Centrality exposure analysis of Flat and Segmented networks connectivity graphs of the two analysed enterprises.}
\vspace{-0.3cm}
\label{fig:outDgr}
\end{figure}
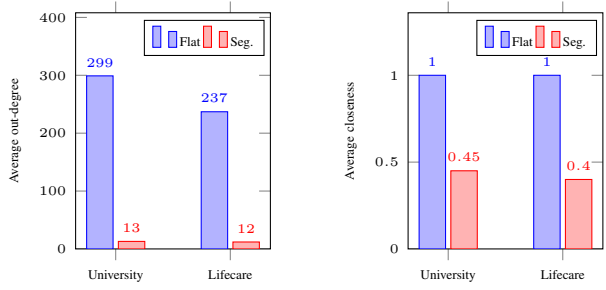

%__________________________________________________________

\begin{figure}[ht]
  \subfloat[Path Length]{
	\begin{minipage}[c][1\width]{
	   0.23\textwidth}
	   \centering
	   \begin{tikzpicture}
         \begin{axis}[  
    ybar,  
    bar width=3 pt,
    enlarge y limits={0.38,upper},
    enlarge x limits=0.15,
    yticklabel style={font=\tiny },
    xticklabel style={font=\tiny },
    ytick style={font=\tiny },
    ylabel={\tiny Paths count},
    xlabel={\tiny Paths length},
    xmin=2, 
    xmax=12,
    ymin=0,
    %ymode=log,
    xtick distance=2,
    nodes near coords,  
    nodes near coords align={horizontal}, 
    nodes near coords style={font=\tiny ,rotate=90, anchor=west},
    width=\textwidth,
    height=0.2\textheight,
    legend pos=north west,
    legend columns=2,
    legend style={nodes={scale=0.5, transform shape}}
    ]  
            \addplot+ coordinates {(2, 359) (4, 5461) (6, 52760) (8, 359281) (10, 1027326) (12, 188253)}; 
            \addplot+ coordinates {(2, 265) (4, 1462) (6, 1538) (8, 190) (10, 0) (12, 0)}; 
            
            \legend{Flat, Seg.}; \end{axis} 
        \end{tikzpicture} 
	\end{minipage}\label{fig:attackPaths}}
 \hfill 
  \subfloat[Betweenness]{
\begin{minipage}[c][1\width]{
	   0.23\textwidth}
	   \centering
	   \begin{tikzpicture}
            \tikzset{
            every pin/.style={fill=yellow!50!white,rectangle,rounded corners=3pt,font=\tiny},
            small dot/.style={fill=black,circle,scale=0.3}
            }
            
            \begin{axis}[
            enlarge y limits={0.35,upper},
            width=1\textwidth,
            height=0.2\textheight,
            ymajorgrids=true,
                grid style=dashed,
                xlabel={\tiny Node ID},
                ylabel={\tiny Node closeness},
            xmin=0, 
            xmax=37,
            ymin=0,
            ymode=log,
            %xtick distance=1,
            legend pos=north east,
            legend columns=2,
            legend style={nodes={scale=1, transform shape}},
            yticklabel style={font=\tiny },
            xticklabel style={font=\tiny },
            no marks,
            thick
            ]
            \addplot table [col sep=comma,trim cells=true,x=ID,y=flat] {graphs/betweeness.csv};
            \addplot table [col sep=comma,trim cells=true,x=ID,y=Seg] {graphs/betweeness.csv};
            
            %\node[small dot,pin=200:{$(5,0.12)$}] at (axis cs:5,0.12) {};
            
            \legend{\tiny Flat , \tiny Seg.  }
            %%\addplot[mark=*] coordinates {(5,0.12)} node[pin=150:{$(5,0.12)$}]{}
            \end{axis}
        \end{tikzpicture}
	\end{minipage}\label{fig:betw}}
	
\caption{Attack graph paths length and betweenness distributions.}
%\caption{Out-degree analysis of the Flat and Segmented networks' attacks graphs of Enterprise A (cf. Tabel~\ref{tab:stats}).}%two analysed enterprises. }
\vspace{-0.3cm}
\label{fig:attackdist}
\end{figure}
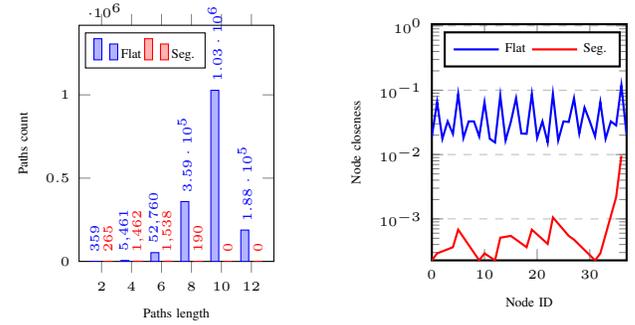

%----------------------

\subsection{Network Robustness Analysis}
\label{subsec:robanalysis}

%To analyse the impact of micro-segmentation on enhancing the robustness of enterprise networks we generate the attack graph of Enterprise A prior to and post the micro-segmentation. We proceed by comparing both attack graphs while relying on the metrics discussed in Section \ref{sec:Risk}.\\
\noindent\textbf{Path-based}: Reducing the number of attack paths is a direct indicator of improving the network robustness reflecting the ability of networks to resist failures or attacks. Assessing the NSP metric representing the count of shortest attack paths, {\it before} and {\it after} micro-segmenting the network of Enterprise A, reveals the efficacy of micro-segmentation to significantly reduce the total number of attack paths by 99\%. 

\begin{table}[h]
\centering
%\caption{Network robustness analysis}
%\vspace{-0.3cm}
\caption{Robustness metrics values of Enterprise A.}
\label{tab:res}
\resizebox{\linewidth}{!}{

\begin{tabular}{|l||c|c|c|}
\hline
\textbf{Metric} & \textbf{Flat} & \textbf{Segmented} & \textbf{Improvement} \\
\hline
\hline
Mis-configurations count & 635  & 221 & 65.2\%  \\
Shortest paths count & 1.633,440  & 3,455 & 99.7\%  \\
Average shortest path length & 5  & 10 & 50\% \\
Min shortest path length & 2  & 2 & 0\%  \\
Min shortest path count & 359  & 265 & 26.2\%  \\
Average out-degree & 28  & 2 & 92.9\%  \\
Maximum out-degree & 35  & 11 & 68.6\%  \\
Average betweenness & 0.04  & 0.0005 & 98.8\%  \\

\hline
\end{tabular}
}
%\vspace{-0.3cm}
\end{table}
We proceed by investigating the distribution of attack paths lengths of Figure \ref{fig:attackPaths}. While the large number of paths with lengths greater than eight in the flat network might give the impression that the flat network is more resilient to attacks as it requires more steps to be performed by the attacker, the total number of attack paths clears the doubt. Unlike the segmented topology, the unrestricted nature of the flat network exposes it to a multitude of direct and indirect potential attack vectors. 

The MSPL metric output, representing the weakest link of the network, is the same for both the flat and micro-segmented networks. However, the count of paths with minimum length, indicating the probability that an attacker exploits the network's weakest link, is reduced by 26\% after micro-segmentation deployment.  

%------------------------------------

\begin{table*}[!t]
\centering
%\vspace{-2cm}
\caption{Attack-graph nodes involving host {\it V}. The first row denotes the attacker acquired privilege on {\it V}. The second row presents the different vulnerabilities of {\it V} where each is identified by the NVD CVE unique identifier \cite{cvss1}. The third row enumerates {\it V} misconfigurations where the insecure connections are defined as hacl(Host1, Host2, Protocol/Service, Port). }%This table %~\ref{tab:app_stats}
%presents the attack graph privilege and configuration nodes involving {\it V}, discussed in Section~\ref{subsec:robanalysis}. 
%Here, the cumulative risk value of acquiring the privilege of executing code on the host {\it V} in the flat network is 0.8 while in the segmented network it is reduced to 0.64. Further analysis of the attack graphs confirms our interpretation of the aggregated risk values.}
%\ik{Elaborate the data in table. Also provide back-references to the section where the reader was referred here. } (cf.\S~\ref{subsec:robanalysis}}
%\label{tab:stats}
\label{tab:app_stats}
\centering
\resizebox{\linewidth }{!}{
\begin{tabular}{|p{1.6cm}||p{9.1cm}|p{6cm}|}
\hline
\textbf{    } & \textbf{Flat Network Attack-graph Nodes} & \textbf{Micro-segmented Network Attack-graph Nodes} \\
\hline
\hline
Privilege & execCode(V,user) 0.8  & execCode(V,user) 0.64 \\
\hline
Vulnerabilities
& 
vulExists (V, CVE-2014-3802, debug interface access software development kit, remote-Exploit, privEscalation), 
vulExists (V, CVE-2017-5715, cortex-a, localExploit, privEscalation), vulExists (V, CVE-2017-5753, cortex-a, localExploit, privEscalation), vulExists (V, CVE-2017-5754, cortex-a, localExploit, privEscalation)
 & 
vulExists (V, CVE-2014-3802, debug interface access software development kit, remote-Exploit, privEscalation), 
vulExists (V, CVE-2017-5715, cortex-a, localExploit, privEscalation), vulExists(V, CVE-2017-5753, cortex-a, localExploit, privEscalation), vulExists (V, CVE-2017-5754, cortex-a, localExploit, privEscalation)
 \\
\hline

Configurations
& 
hasAccount(V victim,  V, user), 
networkServiceInfo(V, debug interface access software development kit, Windows, 2, user), 
hacl (internet, V, Windows, 2), \newline
hacl (V, AL, Windows Microsoft Bulletins, 2),
hacl (V, AG, Windows Microsoft Bulletins, 2),
hacl (V, AU, Windows Microsoft Bulletins, 2),
hacl (V, AU, Windows, 2),
hacl (V, BA, Windows Microsoft Bulletins, 2),
hacl (V, AO, Windows, 2),
hacl (V, HO, Windows Microsoft Bulletins, 2),
hacl (V, KI, Windows Microsoft Bulletins, 2),
hacl (V, KI, Windows, 2),
hacl (V, MA, Windows Microsoft Bulletins, 2),
hacl (v, MI, Windows Microsoft Bulletins, 2),
hacl (V, MI, Windows, 2),
hacl (V, NA, Windows, 2),
hacl (V, NE, Misc, 3),
hacl (V, NE, Windows, 2),
hacl (V, OU, Windows, 2),
hacl (V, OV, Windows Microsoft Bulletins, 2),
hacl (V, PA, Database, 3),
hacl (V, PA, Windows Microsoft Bulletins, 2),
hacl (V, TE, Windows Microsoft Bulletins, 2),
hacl (V, V, Windows, 2),
hacl (V, WI, Windows Microsoft Bulletins, 2),
hacl (V, WI, Windows Microsoft Bulletins, 3),
hacl (TE, V, Windows, 2),
hacl (OV, V, Windows, 2),
hacl (NA, V, Windows, 2),		
hacl (MI, V, Windows, 2),		
hacl (MA, V, Windows, 2),		
hacl (KI, V, Windows, 2),		
hacl (HO, V, Windows, 2),		
hacl (CO, V, Windows, 2),		
hacl (BA, V, Windows, 2),		
hacl (AU, V, Windows, 2),		
hacl (AG, V, Windows, 2),		
hacl (AL, V, Windows, 2),		
hacl (WI, V, Windows, 2)		
& 
hasAccount (V victim, V, user), networkServiceInfo (V, debug interface access software development kit, Windows, 2, user), hacl (internet, V, Windows, 2)\\

\hline
\end{tabular}
}
\vspace{-0.3cm}
\end{table*}
%--------------------------------------
\noindent\textbf{Non-path-based}: Network misconfigurations raise the probability of success in exploiting the assets' vulnerabilities hence increasing the impact of the latter. While micro-segmentation is unable to reduce the vulnerabilities, we claim that modifying the system security configurations influences the likelihood of exploiting the vulnerabilities. We proceed by analysing the count of misconfigurations (CMC) metric. Although the number of network vulnerabilities has not changed as demonstrated by the Nessus scan, we notice that the number of the attack graph root nodes in micro-segmented network is reduced by 65\%, as shown in Table~\ref{tab:res}. Accordingly, we demonstrate that micro-segmentation significantly reduces the attacker accessibility to system vulnerabilities.

The AOD and MOD metrics results, indicating the typical and worst-case number of attacks succeeding a compromised network privilege, reveal the significant impact of micro-segmentation on improving the network robustness. It restricts the attacker's lateral movement and exploration, reducing the average and maximum possible attacks after a network breach by 93\% and 69\%, respectively, as indicated by Table \ref{tab:res}. 

Finally, we analyse the betweenness centrality distribution of the network attack graphs before and after micro-segmentation. Compromised privilege assets with high betweenness centrality have a significant influence on the network robustness by virtue of their control over the paths leading to other privilege nodes, increasing the attacker ability to extend their attained privileges. In fact, micro-segmentation, not only reduced the average betweenness by more than 98\%, it further changed the network topology resulting in a more linear distribution of the betweenness metric. Hence, unlike the jigsaw distribution provided by the flat topology which is characterised by multiple peaks, it evenly distributed the reduced impact of compromising any asset on the network robustness as demonstrated by Figure \ref{fig:betw}.

\noindent\textbf{Common Vulnerability Scoring System (CVSS)}: Next, we analyse the quantitative security of a networked system through the cumulative probability that a network asset is compromised by an attacker. As previously discussed, the probability that an attacker succeeds in obtaining an illegitimate privilege is a function of the severity of a system vulnerability and the accessibility of the vulnerability to the attacker. While the first factor remains intact, after micro-segmentation the second factor is significantly altered as a result of modifying the network topology and limiting the accessibility to the network assets. 

We proceed by calculating the cumulative risk values of the network privilege nodes before and after micro-segmentation following the model in \cite{aggregate}. The distribution of the calculated risk values reveals an average improvement of 20\% after micro-segmentation deployment. While this value represents a significant improvement of the overall enterprise network security risk, it is remarkably lower than the calculated improvements achieved by assessing the network exposure and robustness. This can be attributed to the fact that a major parameter contributing to the risk calculated values, namely the vulnerabilities severity, identified by the Common Vulnerability Scoring System \cite{NVD}, remains unchanged.

To further analyse the latter finding, let us consider the attack graph privilege node \textbf{execCode(V, User)}. It implies that the attacker has gained user privilege to execute code on the victim host {\it V}. As shown in Table \ref{tab:app_stats}, the cumulative risk value of acquiring this privilege in the flat network is found to be 0.8 while in the micro-segmented network it is reduced to 0.64. Further analysis of the attack graphs confirms our interpretation of the aggregated risk values. Both attack graphs have the same vulnerabilities of host {\it V} (four) as represented by the Vulnerabilities row of Table \ref{tab:app_stats}. On the other hand, the flat network has 38 misconfiguration nodes involving {\it V} while the segmented network comprises only three. 

\section{Conclusion}\label{sec:conc}
 
In this paper, we proposed a suite of metrics to analyse the impact of micro-segmentation on improving network security. We leveraged graph features to study the network exposure and robustness against attacks. Hence, the presented work is the {\it first} to formulate {\it objective} graph-features-based metrics for quantifying the effectiveness of micro-segmentation. We rely on real enterprise network data to perform an empirical analysis of the developed suite of metrics. 

The analysis of the formulated metrics before and after the deployment of micro-segmentation proves that the latter is one of the most effective strategies to protect against cyber threats. Comparing and assessing the exposure and robustness metrics reveals an improvement in the range of 60\% -- 90\%.

% The presented work is the first to formulate subjective metrics for quantifying the effectiveness of micro-segmentation and relying on real enterprise network data to perform an empirical analysis of the developed suite of metrics. 
 
% The analysis of the formulated metrics before and post the deployment of micro-segmentation proves that the latter is one of the most effective strategies to protect against cyber threats. Comparing and assessing the exposure and robustness metrics, highlighting possible threats to the enterprise network, reveals an improvement in the range of 60\% -- 90\%.

%It is worth mentioning that the efficacy and efficiency of micro segmentation is not only limited to reducing the network risk. It further contributes in providing a full visibility of the network architecture and configurations to ensure that security policies were applied across all environments in real-time.

\balance

\bibliographystyle{IEEEtranS}
\bibliography{IEEEabrv,NOMS_2022}

%___________________________________________

\end{document}